\documentclass{aa}

\usepackage{graphicx}
\usepackage{natbib}
\usepackage{adjustbox}
\usepackage{txfonts}

\usepackage{xcolor}
\definecolor{yellowgray}{rgb}{0.90, 0.90, 0.2}
\definecolor{bluegray}{rgb}{0.20, 0.60, 0.80}
\definecolor{palered}{rgb}{0.99, 0.40, 0.5}
\definecolor{darkgray}{rgb}{0.35, 0.35, 0.35}
\definecolor{darkgrayb}{rgb}{0.75, 0.75, 0.75}
\definecolor{palegray}{rgb}{0.96, 0.96, 0.96}

\begin{document} 

   \title{Diagnostic capabilities of spectropolarimetric observations for understanding solar phenomena}
  \titlerunning{Diagnostic capabilities of spectropolarimetric observations I}
  \subtitle{I. Zeeman-sensitive photospheric lines}
   \author{C. Quintero Noda \inst{1,2,3,4} \and P. S. Barklem\inst{5} \and R. Gafeira\inst{6,7} \and B. Ruiz Cobo\inst{3,4}    \and M. Collados\inst{3,4} \and  M. Carlsson\inst{1,2} \and \\ V. Mart\'{i}nez Pillet\inst{8}\and D. Orozco Su\'{a}rez\inst{6}  \and H. Uitenbroek\inst{8}  \and Y. Katsukawa\inst{9} 
             }
   \institute{Rosseland Centre for Solar Physics, University of Oslo, P.O. Box 1029 Blindern, N-0315 Oslo, Norway\\
              \email{carlos.quintero@iac.es}     
\and    Institute of Theoretical Astrophysics, University of Oslo, P.O. Box 1029 Blindern, N-0315 Oslo, Norway
\and    Instituto de Astrof\'isica de Canarias, E-38200, La Laguna, Tenerife, Spain.
\and    Departamento de Astrof\'isica, Univ. de La Laguna, La Laguna, Tenerife, E-38205, Spain
\and    Theoretical Astrophysics, Department of Physics and Astronomy, Uppsala University, Box 516, 751 20 Uppsala, Sweden
\and    Instituto de Astrof\'{i}sica de Andaluc\'{i}a (CSIC), Apartado de Correos 3004, E-18080 Granada, Spain
\and    Univ Coimbra, IA, DF, OGAUC, Coimbra, Portugal
\and    National Solar Observatory, University of Colorado Boulder, 3665 Discovery Drive, Boulder, CO 80303, USA
\and    National Astronomical Observatory of Japan, 2-21-1 Osawa, Mitaka, Tokyo 181-8588, Japan\\
             }
   \date{Received  ; accepted  }

 
  \abstract
{Future ground-based telescopes will expand our capabilities for simultaneous multi-line polarimetric observations in a wide range of wavelengths, from the near-ultraviolet to the near-infrared. This creates a strong demand to compare candidate spectral lines to establish a guideline of the lines that are most appropriate for each observation target. We focused in this first work on Zeeman-sensitive photospheric lines in the visible and infrared. We first examined their polarisation signals and response functions using a 1D semi-empirical atmosphere. Then we studied the spatial distribution of the line core intensity and linear and circular polarisation signals using a realistic 3D numerical simulation. We ran inversions of synthetic profiles, and we compared the heights at which we obtain a high correlation between the input and the inferred atmosphere. We also used this opportunity to revisit the atomic information we have on these lines and computed the broadening cross-sections due to collisions with neutral hydrogen atoms for all the studied spectral lines. The results reveal that four spectral lines stand out from the rest for quiet-Sun and network conditions: Fe~{\sc i} 5250.2, 6302, 8468, and 15648~\AA. The first three form higher in the atmosphere, and the last line is mainly sensitive to the atmospheric parameters at the bottom of the photosphere. However, as they reach different heights, we strongly recommend using at least one of the first three candidates together with the Fe~{\sc i} 15648~\AA \ line to optimise our capabilities for inferring the thermal and magnetic properties of the lower atmosphere.}


   \maketitle
%

\section{Introduction}

The decade 2020 represents an excellent opportunity for performing spectropolarimetric observations at high spatial resolution. Ground-based telescopes such as the Daniel K. Inouye Solar Telescope \citep[DKIST,][]{Keil2011} and the European Solar Telescope \citep[EST,][]{Collados2013} will mean a significant leap in the diameters of the primary mirror and consequently, in spatial resolution for work at the diffraction limit. The third flight of the \textsc{Sunrise} balloon-borne telescope \citep{Solanki2010, Barthol2011} is scheduled for 2022, and although it has a lower spatial resolution, it will work at almost atmospheric-free conditions and observe the Sun continuously for several days. In all these missions, polarimetric measurements are the primary target. Most importantly, simultaneous observations of multiple spectral lines will be the baseline of most observing configurations.

A reference compendium of spectral lines optimised for the observing targets of interest is needed. Supporting studies are required that compare these lines at identical conditions and for various physical scenarios. We have recurred to these studies in the past. For example, \cite{Norton2006} provide information based on which we can decide which spectral lines are to be included in the Helioseismic and Magnetic Imager \citep{Schou2012} on board of the Solar Dynamics Observatory \citep{Pesnell2012}. A series of publications started with \cite{Leenaarts2013} to support the Interface Region Imaging Spectrograph \citep{DePontieu2014} mission, and \cite{QuinteroNoda2016, QuinteroNoda2017a, QuinteroNoda2017b, QuinteroNoda2018} to select spectral lines of the \textsc{Sunrise} Chromospheric Infrared spectro-Polarimeter \citep{Katsukawa2020} and the upgraded version of the Imaging Magnetograph eXperiment \citep{MartinezPillet2011} instrument. 

The aim of this work and following papers is to do something similar, but with a much broader scope, focusing on DKIST and EST, as we continue to support \textsc{Sunrise}. The main difference is that this research project needs to cover most of the spectral range from the near-UV to the near-IR, that is, 3800-15000~\AA, which implies a vast number of possible lines that we cannot study in detail. We therefore opt to examine candidate lines that are included in the documentation of these future telescopes. Most of the transitions are well known and have been analysed in the past. Still, one of the critical targets of these telescopes is to observe the most capable lines together, increasing the height resolution of the observations. However, even if we restrict ourselves to the best candidate lines, many spectral lines with different formation heights and capabilities remain from which the atmospheric information may be inferred. We therefore narrow down the target of this work even more to spectral lines that are highly magnetically sensitive and form in the photosphere. After this, we will continue with publications that compare chromospheric lines at identical conditions and non-magnetic spectral lines. Our aims are mainly twofold. On the one hand, we wish to provide a reference  of which spectral regions should be observed in an optimum way for the teams that are responsible for the
instrument development. On the other hand,   we wish to generate a compendium of spectral lines that will help observers of these future missions to tailor their observing proposals based on their science targets.

\begin{center}
\begin{table*}
\normalsize
\caption{Polarisation sensitivity indices $s_Q$ and $s_V$ \citep[][]{Landi2004} for a selection of spectral lines. From left to right, we show the atomic species, the line core wavelength in air, $\log(gf)$, lower and upper level spectroscopic notation, the Land\'{e} factor of the lower and upper level, the effective \textup{first-order} Land\'{e} factor, $s$ and $d$ \citep[see][]{QuinteroNoda2018}, the effective second-order Land\'{e} factor, the line core intensity \citep[extracted from the solar atlas of][]{Delbouille1973}, the normalised line depth $d_c$, and the circular and linear polarisation sensitivity index. We define $\lambda_{\rm ref}$=5000~\AA ,\ and the continuum corresponds to $I_c=10000$ (in arbitrary units, AU) for all the lines.} \label{atomic_info}  
\begin{adjustbox}{width=0.96\textwidth}
  \bgroup
\def\arraystretch{1.25}
\begin{tabular}{lcccccccccccccc}
        \hline
Atom & $\lambda$ [\AA] & $\log(gf)$ &   $L_1$ & $L_2$  & $g_1$ & $g_2$ & $\bar{g}$ & $s$ & $d$ & $\bar{G}$  & $I(\lambda_0)$ [AU]  & $d_c$ & $s_Q$ & $s_V$  \\
        \hline
Fe~{\sc i}   & 5247.0501   & -4.946 & ${}^5D_{2}$       & ${}^7D^{\rm o}_{3}$   &  1.50  &  1.75 & 2.00 & 18.00  & -6.00  & 3.98  & 2808 & 0.719 & 1.509 &      3.152   \\       
Fe~{\sc i}   & 5250.2086   & -4.938 & ${}^5D_{0}$       & ${}^7D^{\rm o}_{1}$   &  0.00  &  3.00 & 3.00 & 2.00  & -2.00  & 9.00  & 2841 & 0.715 & 2.252 & 7.094   \\
Fe~{\sc i}   & 5250.6456   & -2.181 & ${}^5P_{2}$       & ${}^5P^{\rm o}_{3}$   &  1.83  &  1.67 & 1.50 & 18.00 & -6.00  & 2.24  & 2028 & 0.796 & 1.254 & 1.967   \\
Fe~{\sc i}   & 6173.3352   & -2.880 & ${}^5P_{1}$       & ${}^5D^{\rm o}_{0}$   &  2.50  &  0.00 & 2.50 & 2.00  & 2.00   & 6.25  & 3738 & 0.626 & 1.933 & 5.966    \\
Fe~{\sc i}   & 6301.5008   & -0.718 & ${}^5P^{\rm o}_{2}$   & ${}^5D_{2}$       &  1.83  &  1.50 & 1.67 & 12.00 & 0.00   & 2.52  & 2780 & 0.722 & 1.520 & 2.890   \\
Fe~{\sc i}   & 6302.4932   & -1.131* & ${}^5P^{\rm o}_{1}$   & ${}^5D_{0}$      &  2.50  &  0.00 & 2.50 & 2.00  & 2.00   & 6.25  & 3443 & 0.656 & 2.066 & 6.511   \\
Ni~{\sc i}   & 6767.7700   & -2.170 & ${}^1S_{0}$   & ${}^3P^{\rm o}_{1}$      &  0.00  &  1.50 & 1.50 & 2.00  & -2.00   & 2.25  & 3630 & 0.637 & 1.293 & 2.626   \\
Fe~{\sc i}   & 6820.3715   & -1.290 & ${}^5P^{\rm o}_{1}$   & ${}^5P_{2}$      &  2.50  &  1.83 & 1.50 & 8.00  & -4.00   & 2.18  & 6453 & 0.355 & 0.726 & 1.439   \\
Fe~{\sc i}   & 6842.6854   & -1.290 & ${}^5P^{\rm o}_{1}$   & ${}^5P_{1}$      &  2.50  &  2.50 & 2.50 & 4.00  & 0.00   & 6.25  & 6589 & 0.341 & 1.167 & 3.992   \\
Fe~{\sc i}   & 8468.4069   & -2.072 & ${}^5P_{1}$       & ${}^5P^{\rm o}_{1}$   &  2.50  &  2.50 & 2.50 & 4.00  & 0.00   & 6.25  & 3643 & 0.636 & 2.691 & 11.40   \\
Fe~{\sc i}   & 8514.0716   & -2.229 & ${}^5P_{2}$       & ${}^5P^{\rm o}_{2}$   &  1.83  &  1.83 & 1.83 & 12.00  & 0.00   & 3.36  & 3813 & 0.619 & 1.928 & 6.027   \\
Fe~{\sc i}   & 10783.051   & -2.571* & ${}^3P_{0}$       & ${}^3P^{\rm o}_{1}$   &  0.00  &  1.50 & 1.50 & 2.00  & -2.00   & 2.25  & 6877 & 0.312 & 1.010 & 3.268   \\
Si~{\sc i}   & 10786.856   & -0.365 & ${}^3P^{\rm o}_{1}$       & ${}^3P_{0}$   &  1.50  &  0.00 & 1.50 & 2.00  & 2.00   & 2.25  & 3906 & 0.609 & 1.972 & 6.382   \\
Fe~{\sc i}   & 15534.246   & -0.585* & ${}^5D_{1}$       & ${}^5P^{\rm o}_{2}$   &  1.50  &  1.83 & 2.00 & 8.00  & -4.00   & 3.98  & 6606 & 0.339 & 2.109 & 13.04   \\
Fe~{\sc i}   & 15542.083   & -0.457* & ${}^5D_{1}$       & ${}^5D^{\rm o}_{0}$   &  1.50  &  0.00 & 1.50 & 2.00  & 2.00   & 2.25  & 7157 & 0.284 & 1.326 & 6.181   \\
Fe~{\sc i}   & 15648.514   & -0.714* & ${}^7D_{1}$       & ${}^7D^{\rm o}_{1}$   &  3.00  &  3.00 & 3.00 & 4.00  & 0.00   & 9.00  & 7003 & 0.300 & 2.814 & 26.42   \\
        \hline
  \end{tabular}
  \egroup
\end{adjustbox}  
\end{table*}
\end{center}

\section{Method}\label{method}

\subsection{Synthesis of the Stokes profiles}

We make use of the code called Stokes inversion based on response functions \citep[ SIR;][]{RuizCobo1992} to synthese the full Stokes vector. The code assumes local thermodynamic equilibrium (LTE).

\begin{figure}
\begin{center} 
 \includegraphics[trim=0 0 0 0,width=8.5cm]{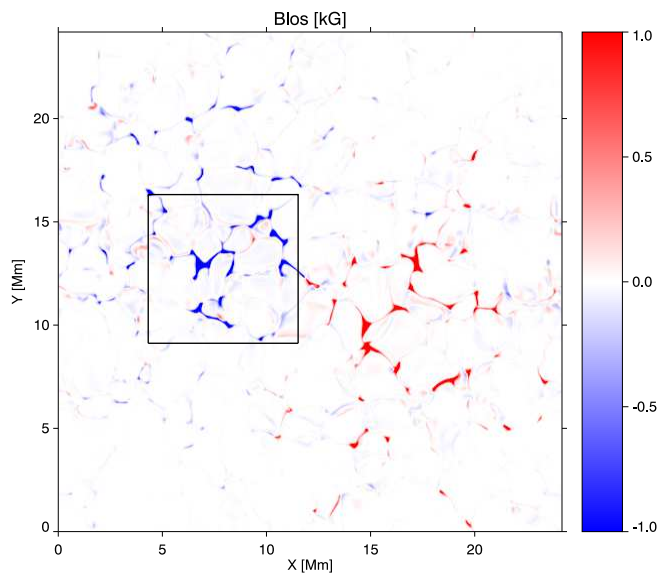}
 \caption{Longitudinal field strength at the 5000~\AA \ continuum optical depth $\log \tau=0$ in the snapshot 385 from the {\sc bifrost} enhanced network simulation.}
 \label{context}
 \end{center}
\end{figure}

We used two types of atmospheric models, 1D semi-empirical atmospheres, and 3D realistic numerical simulations. In the first case, we employed the Harvard-Smithsonian Reference Atmosphere \citep[HSRA,][]{Gingerich1971} to compute the dependence of the polarisation signals on the field strength and the response functions (RF) \citep[e.g.][]{Landi1977} to perturbations of the atmospheric parameters. In the second case, we used the snapshot 385 of the enhanced network simulation described in \cite{Carlsson2016} and developed with the Bifrost code \citep{Gudiksen2011}. The snapshot covers a surface of $24\times24$~Mm$^2$ with a pixel size of 48~km, while the vertical domain extends from 2.4~Mm below to 14.4~Mm above the average optical depth unity at $\lambda=5000$~\AA. Although the simulated scenario is simpler than recent numerical simulations such as were presented in \cite{Hansteen2017, Hansteen2019}, and \cite{Cheung2019}, it contains a configuration with strong network patches and quiet areas. These structures allow ascertaining the capabilities of the spectral lines for quiet-Sun spectropolarimetry. Moreover, the simulation is publicly available\footnote{\url{http://sdc.uio.no/search/simulations}}, so that our work can be extended with additional lines that can be compared with our results.

For both studies we assumed disk centre observations, that is, $\mu=1$, where $\mu=\cos(\theta)$ and $\theta$ is the heliocentric angle. We used the abundance values of the different atomic species given in \cite{Asplund2009}. We defined a spectral sampling of 10~m\AA. We did not degrade the Stokes profiles with any macroturbulence correction or by adding the effect of noise or spatial resolution. Finally, we did not include any microturbulence enhancement for the HSRA atmosphere and the enhanced network simulation. Previous studies \citep{Leenaarts2009} mentioned the lack of small-scale velocities in the Bifrost simulation that should be accounted for because otherwise, the polarisation signals would be artificially enhanced \citep[see][]{delaCruzRodriguez2012}. See also the detailed discussion presented in \cite{Carlin2017} for a more in-depth analysis of the impact of a microturbulence enhancement on the polarisation signals for this simulation. However, we are more interested here in the relative differences between the spectral lines of interest than in comparing with actual observations. 

\subsection{Spectral lines}

We performed a thorough selection of spectral lines based on different sources, mainly the EST, DKIST, and \textsc{Sunrise} science requirement documents, and Table 1 in \cite{Smitha2017}. We tried to cover as many candidate lines for photospheric Zeeman spectropolarimetry as possible, but we may have missed some of them. We are therefore open to suggestions for future publications.

Table~\ref{atomic_info} contains the spectroscopic information of the spectral lines in which we are interested. We extracted this information from the NIST database \citep{nist2018} as the primary source. However, there are cases where the $\log$~$gf$ is not specified, therefore we used the R.~Kurucz \citep{Kurucz1995} database instead (see asterisks). We followed the same steps as \cite{QuinteroNoda2018} to compute the sensitivity indices to linear and circular polarisation as explained in the monograph of \cite{Landi2004}. The advantage of using these effective values is that we also considered the spectral line wavelength and line core intensity. Therefore they provide a more accurate estimation of the sensitivity of the spectral lines to the Zeeman effect \citep[see e.g. the study presented in][]{MartinezGonzalez2008}.  \cite{QuinteroNoda2018} described how each of the necessary elements was computed to obtain the first-order $\bar{g}$ and second-order $\bar{G}$ Land\'{e} factors, as well as the sensitivity indices $s_Q$ and $s_V$. We therefore refer to this work for more information and to \cite{Landi2004} for a complete description. The solar atlas used for the computation was observed by \cite{Delbouille1973} and provided by the BASS200 solar archive\footnote{\url{http://bass2000.obspm.fr/solar_spect.php}}. The continuum was estimated locally (in wavelength), and the intensities were normalized so that the continuum was 100\% (value 10000 arbitrary units, AU). In our case, $I_c=10000$ AU for all the spectral lines.

We summarise some context information and publications that have used these spectral lines to explain why we considered them here. We only describe here the lines in detail later below, while the so-called auxiliary lines are included in Table~\ref{atomic_info} for completeness and are justified in the next section.

We start with the Fe~{\sc i}  5250.2~\AA \ line, which has been widely used for polarimetric observations since the early 1970s, for instance, by \cite{Stenflo1973}, because it has an effective Land\'{e} factor of 3 and also strongly depends on the temperature. Modern instrumentation such as the Imaging Magnetograph eXperiment \citep{MartinezPillet2011} on board the \textsc{Sunrise} balloon-borne telescope \citep{Solanki2010, Barthol2011} also observes it to further improve our knowledge of the magnetic field in the lower atmosphere \citep[see e.g. the recent review of][]{BellotRubio2019}. The next photospheric candidate is the Fe~{\sc i}  6173~\AA \ spectral line, which  has become one of the best transitions for space magnetographs because it has a clean continuum and no blends \citep{Norton2006}. It is observed routinely by the Helioseismic and Magnetic Imager \citep{Schou2012} on board the Solar Dynamics Observatory \citep{Pesnell2012} and will also be observed by the Polarimetric and Helioseismic Imager \citep{Solanki2019} on board the Solar Orbiter mission \citep{Muller2013}. The following spectral line is the Fe~{\sc i} 6302.5~\AA, which was extensively recorded by the Advanced Stokes Polarimeter \citep{Elmore1992} in the 1990s and was later the main spectral line for analysing the solar photosphere at high spatial resolution from space through the observations made by the Spectropolarimeter \citep{Lites2013} of the Solar optical telescope \citep{Tsuneta2008} on board the Hinode spacecraft \citep{Kosugi2007}.  The next transition we studied is the Ni~{\sc i} 6768~\AA \ line, one of the most famous spectral lines for helioseismology and the analysis of large-scale photospheric magnetism. It was routinely observed by the Michelson Doppler Imager \citep{Scherrer1995} on board the SoHo mission \citep{Domingo1995}, and it is recorded daily by the Global Oscillation Network Group project \citep{Harvey1996} to study the solar internal structure through helioseismology. The Fe~{\sc i} 6842~\AA \ spectral line has a high sensitivity to the magnetic field with $\bar{g}=2.5$, and it also forms in the low photosphere \citep[e.g. ][]{Smitha2017}. The next transition is the Fe~{\sc i} 8468~\AA , which is located relatively close to the Ca~{\sc ii} 8498 and 8542~\AA \ chromospheric lines. Hence, it is part of the critical spectral lines of the \textsc{Sunrise} Chromospheric Infrared spectro-Polarimeter \citep{Katsukawa2020} that will simultaneously observe the 770~nm region \citep{QuinteroNoda2017b} and the 850~nm window described in \cite{QuinteroNoda2017a}. The line is strong, and has an effective Land\'{e} factor of 2.5 as well. The Fe~{\sc i} 10783~\AA \ line has a lower magnetic field sensitivity,  $\bar{g}=1.5$, but is a target line for future missions because it is well suited for sunspot observations as it does not show indications of blends in the umbra \citep{Balthasar2008}. The next transition on the list, that is, Fe~{\sc i} 15534~\AA, with $\bar{g}=2.0,$ is a good candidate for line-ratio techniques \citep{Smitha2017} and can be observed at present, for instance, with the GRIS spectrograph \citep{Collados2012} installed on the Gregor telescope \citep{Schmidt2012}. Finally, GRIS as well as its predecessor \citep[the Tenerife Infrared Polarimeter,][]{Collados2007} has routinely observed the Fe~{\sc i} 15648~\AA \ line, the last selected transition for this work. This line is one of the few candidates that has a Land\'{e} factor of 3.0, and it is one of the lines of excellence for quiet-Sun observations \citep[see e.g.][for more information]{BellotRubio2019}.

\begin{figure*}
\begin{center} 
 \includegraphics[trim=0 0 0 0,width=18.2cm]{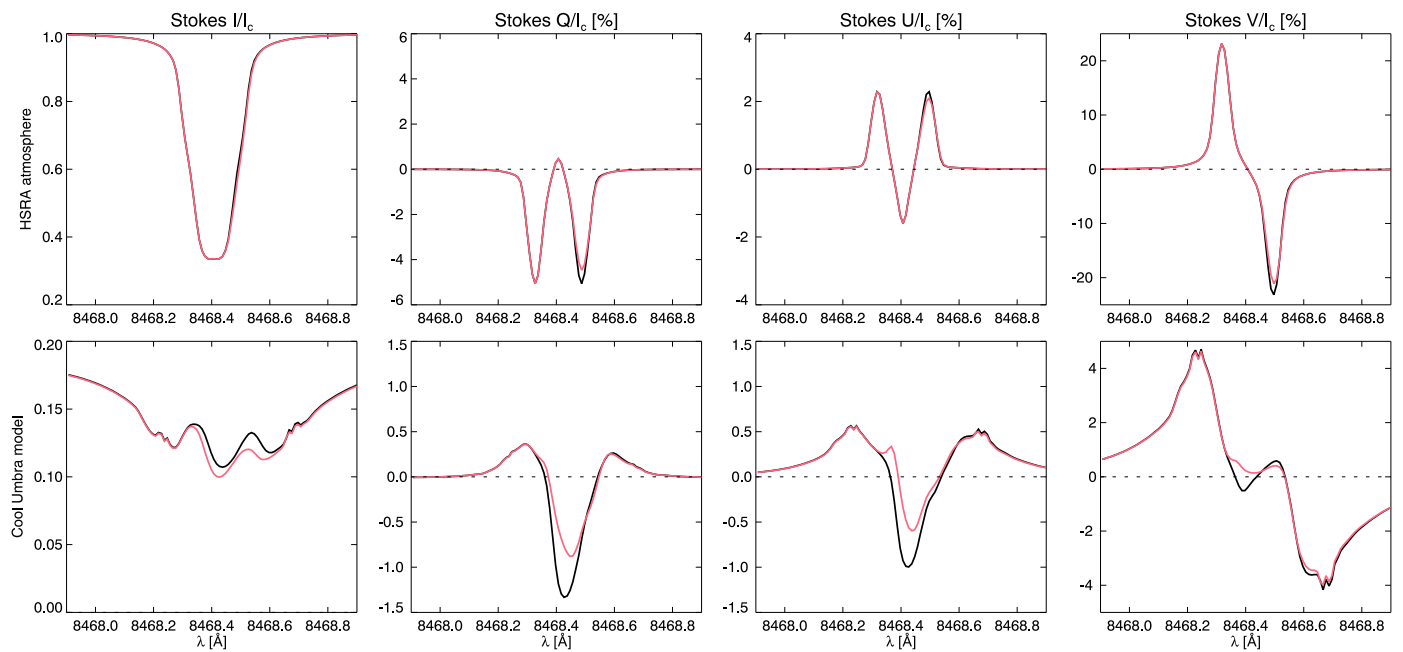}
 \caption{Comparison of the synthesis of the Fe~{\sc i} 8468.4069 \AA \ alone (black) and the synthesis of this line blended with Ti~{\sc i} 8468.4700 \AA \  (red). The upper row corresponds to the results using the HSRA semi-empirical atmosphere with a magnetic field of 500~G, and 45 and 70 degrees of magnetic inclination and azimuth, respectively. The lower row displays the synthetic profiles from the cool-umbra model atmosphere.}
 \label{LTE_8468}
 \end{center}
\end{figure*}

\subsection{Auxiliary spectral lines}

A few of the spectral lines included in Table~\ref{atomic_info} are not strictly the best candidates for Zeeman polarimetry. Their effective Land{\'e} factor is not as high as that of the other lines, for example, or they do not present particular properties in specific magnetic configurations. However, they are similar to those that are very sensitive to the magnetic field and were used in the past to estimate, for instance, the field strength of quiet-Sun observations \citep{Stenflo1973}. Stenflo (1973) computed the ratio of the Fe~{\sc i} 5247 and 5250.2 \AA \ lines because the atomic levels that produced their transitions are identical, except for the total angular momentum values. This technique allowed them to infer the field strength from the Stokes $V$ amplitude ratio because the difference between the two lines should be the Land{\'e} factor.  It was therefore customary to observe them together, and we expect that some observers will do the same with future facilities. We therefore also included the Fe~{\sc i} 8468 and 8515 \AA \ as they represent a similar case. Furthermore, the 8515~\AA \ line lies between the two Ca~{\sc ii} 8498 and 8542~\AA \ lines and would represent an excellent photospheric complement to these two chromospheric lines. We also added the two Fe~{\sc i} 10783 and Si~{\sc i} 10786 \AA \ lines as they are close enough to be observed together and can complement each other. Finally, we list in Table~\ref{cross} additional lines that fall close to the target spectral lines of this work and are usually observed together, for instance, Fe~{\sc i} 5250.6, 6301.5, or 15652~\AA.

\begin{table*}
\normalsize
\caption {Atomic information used to compute the collisional broadening parameters. From left to right, we list the atom species, transition wavelength, the dominant configuration and term designation for the lower and upper levels, the energy of the lower level, the series limit energy for that level, the energy of the upper level, the series limit energy for that level, the orbital angular momentum quantum numbers of the optical electron in the lower and upper levels, the total angular momentum quantum numbers of the lower and upper levels, and finally the calculated line broadening cross-section $\sigma$ due to hydrogen atom impact at a collision velocity of $10^4$~m/s and corresponding velocity dependence parameter $\alpha$. The units for the wavelength are \AA, the energy is listed in cm$^{-1}$, and atomic units for the cross-section $\sigma$, respectively. The velocity parameter $\alpha$ is dimensionless.} \label{cross} 
\begin{adjustbox}{width=1.00\textwidth}
  \bgroup
\def\arraystretch{1.25}
\begin{tabular}{lcccccccccccccc}
        \hline
Atom & $\lambda$ [\AA] &        Lower configuration and term & Upper configuration and term & $E_{lower}$ [$cm^{-1}$] &  $E^{limit}_{lower}$ [$cm^{-1}$] & $E_{upper}$ [$cm^{-1}$] &   $E^{limit}_{upper}$ [$cm^{-1}$] & $l_{low}$ & $l_{upp}$ & $J_{low}$ & $J_{upp}$ & $\sigma$ [a.u.] & $\alpha$ \\
        \hline  
Fe~{\sc i}   & 5247.0501   & $3d^{6}4s^{2}$      a ${}^5$D & $3d^6$(${}^5$D)$4s4p$(${}^3$P$^{\rm o})$ z ${}^7$D$^{\rm o}$ & 704.007 & 63737.704 & 19757.032 & 63737.704 & 0 &  1 &  2 &   3 & 205 &  0.255 \\             
Fe~{\sc i}   & 5250.2086   & $3d^{6}4s^{2}$      a ${}^5$D & $3d^6$(${}^5$D)$4s4p$(${}^3$P$^{\rm o})$ z ${}^7$D$^{\rm o}$ & 978.074 & 63737.704 & 20019.635 & 63737.704 & 0 &  1 &  0 &   1 & 206 & 0.254 \\
Fe~{\sc i}   & 5250.6456   & $3d^7({}^4$P)$4s$   a ${}^5$P & $3d^6$(${}^5$D)$4s4p$(${}^1$P$^{\rm o}$) y ${}^5$P$^{\rm o}$ & 17726.988 & 76933.840 & 36766.966 & 63578.161 & 0 &  1 &  2 &   3 & 343 & 0.266 \\
Fe~{\sc i}   & 6173.3352   & $3d^7({}^4$P)$4s$   a ${}^5$P & $3d^7$(${}^4$F)$4p$  y ${}^5$D$^{\rm o}$ & 17927.382 & 76933.840 & 34121.603 & 65738.010 & 0 &  1 &  1 &   0 & 279 &  0.245 \\
Fe~{\sc i}   & 6301.5008   & $3d^6$(${}^5$D)$4s4p$(${}^3$P$^{\rm o}$) z ${}^5$P$^{\rm o}$  & $3d^6$(${}^5$D)$4s$(${}^6$D)5$s$ e ${}^5$D & 29469.024 & 63737.704 & 45333.875 & 63737.704 & 1 &  0 &  2 &   2 & 834 &  0.242 \\
Fe~{\sc i}   & 6302.4932   & $3d^6$(${}^5$D)$4s4p$(${}^3$P$^{\rm o}$)    z ${}^5$P$^{\rm o}$ & $3d^6$(${}^5$D)$4s$(${}^6$D)5$s$ e ${}^5$D & 29732.736 & 63737.704 & 45595.086 & 63737.704 & 1 &  0 &  1 &   0 & 850 &  0.239 \\
Ni~{\sc i}   & 6767.7700   & $3d^{10}$ ${}^1$S & $3d^9$(${}^2$D)$4p$ ${}^3$P$^{\rm o}$ & 14728.840 & 61619.77 & 29500.674 & 61619.77 & 2 &  1 &  0 &   1 & 301 &  0.241 \\
Fe~{\sc i}   & 6820.3715   & $3d^6$(${}^5$D)$4s4p$(${}^1$P$^{\rm o}$)    y ${}^5$P$^{\rm o}$ & $3d^6$(${}^5$D)$4s$(${}^6$D)4$d$ e ${}^5$P & 37409.555 & 63737.704 & 52067.469 & 63737.704 & 1 &  2 &  1 &   2 & 897 &  0.282 \\
Fe~{\sc i}   & 6842.6854   & $3d^6$(${}^5$D)$4s4p$(${}^1$P$^{\rm o}$) y ${}^5$P$^{\rm o}$ & $3d^6$(${}^5$D)$4s$(${}^6$D)4$d$ e ${}^5$P & 37409.555 & 63737.704 & 52019.669  & 63737.704 & 1 &  2 &  1 &   1 & 891 &  0.281 \\
Fe~{\sc i}   & 8468.4069   & $3d^7({}^4$P)$4s$ a ${}^5$P & $3d^6$(${}^5$D)$4s4p$(${}^3$P$^{\rm o}$) z ${}^5$P$^{\rm o}$& 17927.382 & 76933.840 & 29732.736 & 63737.704 & 0 &  1 &  1 &   1 & 259  & 0.245 \\
Ti~{\sc i}   & 8468.4700   &   $3d^3({}^2$G)$4s$         a ${}^3$G& 3$d^2$(${}^1$D)4s4p(${}^3$P$^{\rm o}$) x ${}^3$F$^{\rm o}$        & 15220.393 & 63912.060 & 27025.659 & 63578.161 & 0 &  1 &  5 &   4 & 268 &   0.252 \\
Fe~{\sc i}   & 8514.0716   & $3d^7({}^4$P)$4s$ a ${}^5$P& $3d^6$(${}^5$D)$4s4p$(${}^3$P$^{\rm o}$) z ${}^5$P$^{\rm o}$ & 17726.988  & 76933.840 & 29469.024  & 63737.704 & 0 &  1 &  2 &   2 & 257  & 0.246 \\
Fe~{\sc i}   & 10783.051   & $3d^7({}^2$P)$4s$ c ${}^3$P& $3d^6$(${}^5$D)$4s4p$(${}^3$P$^{\rm o}$) z ${}^3$P$^{\rm o}$& 25091.599  & 81857.423 & 34362.873  & 63737.704 & 0 &  1 &  0 &   1 & 306 &  0.245 \\
Si~{\sc i}   & 10786.856   & $3s^23p4p$ ${}^3$D 2  & $3s^23p4d$ ${}^3$F$^{\rm o}$ & 48102.323   & 66035.000  & 57372.297  & 66035.000 & 1 &  2 &  2 &   2 & 1341 & 0.302 \\
Fe~{\sc i}   & 15534.246   & $3d^6$(${}^5$D)$4s$ (${}^6$D)$5s$ e ${}^5$D& $3d^6$(${}^5$D)$4s$ (${}^6$D)$5p$       u ${}^5$P$^{\rm o}$ & 45509.152  & 63737.704 & 51944.784          & 63737.704 & 0 &  1 &  1 &   2 & 1180 &  0.237 \\
Fe~{\sc i}   & 15542.083   & $3d^6$(${}^5$D)$4s$ (${}^6$D)$5s$ e ${}^5$D & $3d^6$(${}^5$D)$4s$ (${}^6$D)$5p$ t ${}^5$D$^{\rm o}$ & 45509.152  & 63737.704 & 51941.540  & 63737.704 & 0 &  1 &  1 &   0 & 1179 &  0.237 \\
Fe~{\sc i}   & 15648.514   & $3d^6$(${}^5$D)$4s$ (${}^6$D)$5s$ e ${}^7$D & $3d^6$(${}^5$D)$4s$ (${}^6$D)$5p$ n ${}^7$D$^{\rm o}$ & 43763.980 & 63737.704 & 50152.619 & 63737.704 & 0 &  1 &  1 &   1 & 974 &  0.229 \\
Fe~{\sc i}   & 15652.873   & $3d^6$(${}^5$D)$4s$ (${}^6$D)$4d$ f ${}^7$D & $3d^64s$(${}^6$D$_{9/2}$)$4f$ (${}^2[7/2]^{\rm o}$) & 50377.908 & 63737.704 & 56764.767 & 63737.704 & 2 &  3 &  5 &   4 & 1428 &  0.330 \\
        \hline
  \end{tabular}
  \egroup
\end{adjustbox}
\vspace{0.1cm}      
\end{table*}

\subsection{Fe~{\sc i} 8468.4069 \AA \ and Ti~{\sc i} 8468.4700 \AA}

In previous works, we presented the Fe~{\sc i} 8468.4069 \AA \ as one of the most capable photospheric spectral lines (e.g. \cite{QuinteroNoda2017a}). It should therefore be included in this study as well. However, when we presented the preliminary results of this work at the National Solar Observatory, we received some helpful feedback from Alexandra Tritschler. She mentioned that there is a Ti~{\sc i} transition close to the iron line at 8468.4700~\AA \ that might modify its profile shape, mainly for cool atmospheres where titanium lines are usually enhanced \citep[see e.g. the review of][]{Solanki2003}. We therefore decided to quickly compare the results obtained by synthesising only the iron transition or both spectral lines in this publication. The results are plotted in Figure~\ref{LTE_8468}, where we used the HSRA atmosphere adding a constant magnetic field (top) and the cool umbra model presented in \cite{Collados1994} (bottom). When the Ti~{\sc i} line is included, the Stokes profiles in the HSRA atmosphere are slightly modified. The perturbation is mainly in the red wing of the iron line as the titanium transition falls at 63.1~m\AA \ from its line core. In the case of the cool model, the differences are more significant, and the perturbation appears to be closer to the centre of the iron line. From now on, we therefore need to include the titanium line when synthesising or inverting the Fe~{\sc i} 8468.4069~\AA \ transition (as we do in this publication).

\subsection{Collisions with neutral hydrogen}

The broadening of the spectral lines by collisions with neutral hydrogen atoms is computed using the Anstee, Barklem and O'Mara (ABO) theory \citep[e.g.][]{Anstee1995, Barklem1997}, specifically the abo-cross calculator code \citep{Barklem2015}, which interpolates in pre-computed tables of line broadening parameters. A description of how to use the code can be found in \cite{Barklem1998}.  For clarity, we describe two examples below, one that is straightforward, and another that involves a two-electron excitation.  We show in Table~\ref{cross} the data we used for the transitions of interest and the resulting calculated values of the line broadening parameters, the cross-section $\sigma$ at a collision velocity of $10^4$~m/s, and the velocity parameter $\alpha$.  See \cite{Barklem1998} for example for an explanation of these parameters and the formula for calculating the line widths.  We include the target lines of this work and the additional transitions that fall close to the spectral lines of interest. 

We start with the straightforward example of the Fe~{\sc i} 5250.2~\AA \ transition.  We extracted the basic information from the NIST database, including the energy of each level ($E_{lower}$ and $E_{upper}$) and the orbital angular momentum of the optical electron ($l$). The total angular momentum ($J$) was not used in the calculation, but was extracted for identification purposes. The remaining input parameters were the series limits for the configurations of each level $E_{limit}$, which defines the binding energy of the optical electron \citep[see][for a detailed description and examples]{Barklem1998}.  For the two levels involved in this transition, the parent configuration is $3d^{6}4s$, corresponding to the ground configuration and level of Fe~{\sc ii}. Thus, $E_{limit}$ is simply the ionisation energy of Fe~{\sc i}, that is, 63737.704~cm$^{-1}$. In the case of the Fe~{\sc i} 5250.6 \AA \ transition, we have the same situation for the upper level: After removing the optical electron, the parent configuration is again $3d^{6}4s$, and so the series limit energy is also $E_{limit}=63737.704$~cm$^{-1}$. However, in the lower level of the transition, the parent $3d^7\,{}^4$P corresponds to an excited level of Fe~{\sc ii}; that is, at the Fe~{\sc i} level, two electrons are excited, the optical electron and a core electron.  The binding energy of the optical electron must account for this excited core. Thus the series limit energy should be the sum of the ionisation energy for Fe~{\sc i} and the energy of this level of Fe~{\sc ii} (relative to the ground state of Fe~{\sc ii}) as this is the level remaining if the optical electron is removed, thus defining the binding energy.  NIST gives a term energy for $3d^7\,{}^4$P of 13196.137~cm$^{-1}$, and thus the series energy of the lower level is $E_{limit}=63737.704+13196.137=76933.840$~cm$^{-1}$.  We chose to use the term value because selecting a parent with a specific $J$ or a mixture of $J$ values is not justified by the theory, which ignores spin and angular momentum coupling with the core.  Furthermore, this would change the calculation only very little, which is a negligible error relative to the overall accuracy of the theory.   

Some of the lines fell just outside the bounds of the tables included in the code; for these transitions, specific calculations were performed. The case of Ni~{\sc i} 6768~\AA \ requires special mention. The lower state has very tightly bound equivalent valence electrons in $d$ orbitals, the wavefunction for which cannot be calculated with the Coulomb wavefunction method employed in the ABO theory.  The broadening of a spectral line is usually dominated by the upper level, with some interference effects due to the lower level.  To calculate the broadening of this line, including some estimate of the interference due to the lower level, we therefore assumed that the lower level was spherically symmetric ($s$ state), allowing ABO codes to be employed, and the value we calculated is given in Table~2. Ignoring the lower state completely leads to a change (increase) of about 7~\%, indicating that the lower level indeed contributes only little to the broadening through interference effects.

As a final note, in the case of the SIR code, the cross-section $\sigma$ needs to be provided in cgs units, that is, cm$^2$. We therefore need to multiply the $\sigma$ values given in the table by $a_{0}^2=2.80028\times 10^{-17}$~cm$^2$.

\begin{figure}
\begin{center} 
 \includegraphics[trim=0 0 0 0,width=8.0cm]{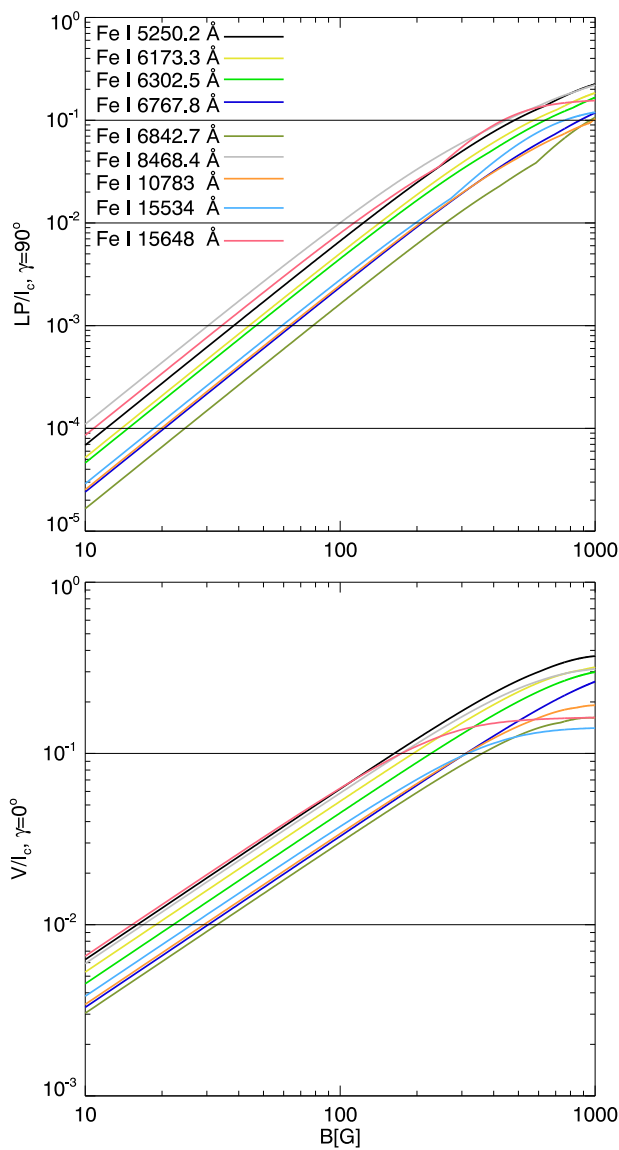}
 \caption{Maximum linear ($LP=max(\sqrt{Q^2+U^2})$) and circular ($V=max(|V|)$) polarisation signals for the selected nine most Zeeman-sensitive photospheric lines (see legend in the top panel). We use the HSRA atmosphere and vary the field strength from 10 to 1000 G (see abscissas). The upper panel shows the results when the inclination is 90 degrees (horizontal magnetic field with respect to the solar surface), and the lower panel displays the results for a magnetic field that is parallel to the line of sight, i.e. an inclination of 0 degrees. The azimuth is always equal to 70 degrees.}
 \label{LTE_pol}
 \end{center}
\end{figure}

\begin{figure*}
\begin{center} 
 \includegraphics[trim=0 0 0 0,width=18.2cm]{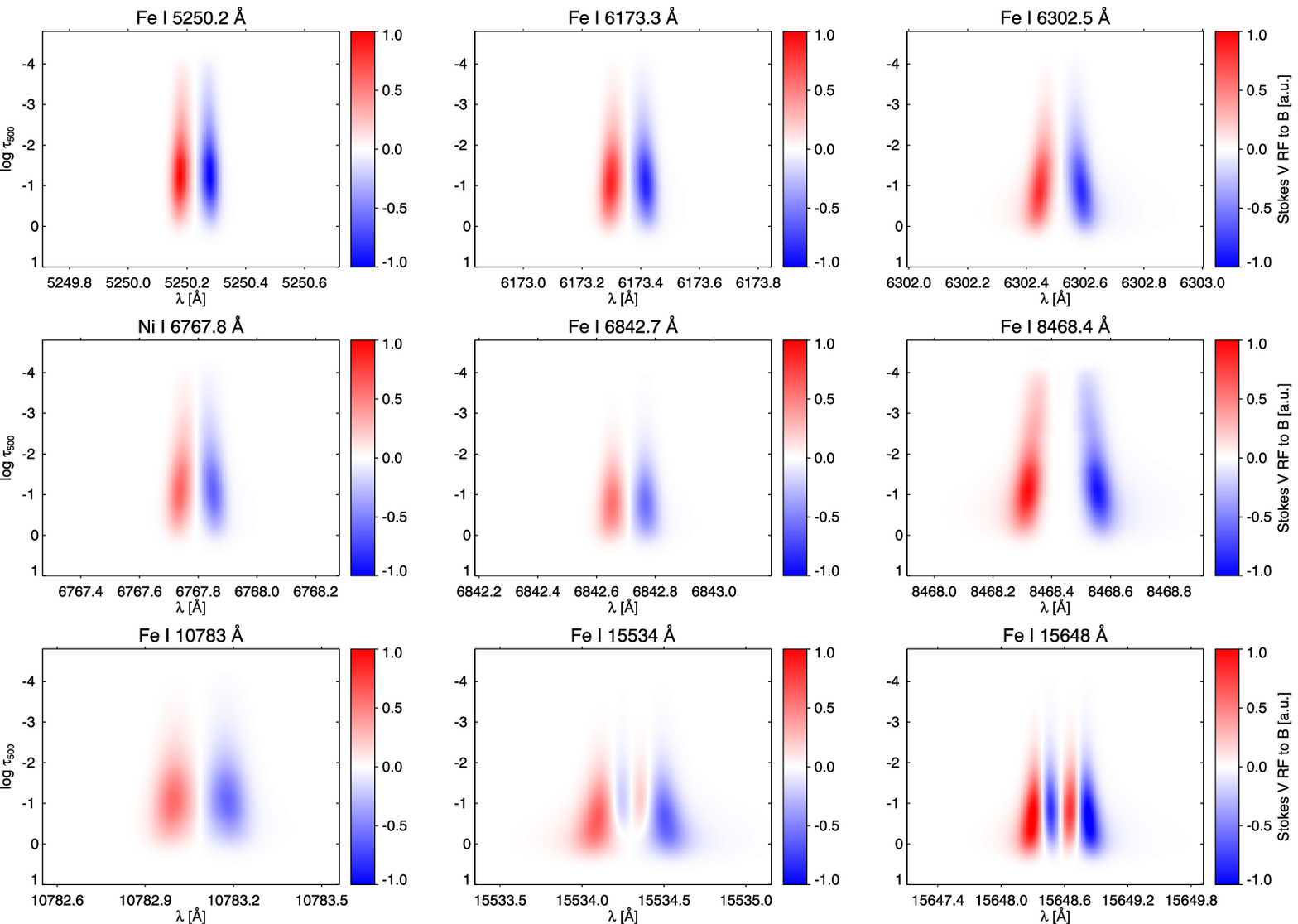}
 \caption{Two-dimensional plot of the Stokes $V$ RF to changes in the field strength. Each panel displays the results for one of the selected spectral lines. White areas designate regions of no sensitivity to changes in the atmospheric parameters, and red and blue indicate opposite signs of the RF.  All panels are normalised to the maximum of the Fe~{\sc i} 5250.2~\AA \ Stokes $V$ RF.}
 \label{LTE_2DRF}
 \end{center}
\end{figure*}

\section{Spectral line analysis based on 1D semi-empirical atmospheres}\label{1D}

\subsection{Maximum polarisation signals}

To compute the maximum polarisation signals, we started with the HSRA atmosphere, including a magnetic field constant with height. Then we modified its field strength using 1~G steps from 10 to 1000~G. As we planned to estimate the transversal and longitudinal signals separately, we performed the computation with a magnetic field inclination ($\gamma$) of 90~degrees first and then with $\gamma=0$~degrees. In both cases, the magnetic field azimuth is equal to 70~degrees.
 
The top panel of Figure~\ref{LTE_pol} displays the total linear polarisation signals (computed as $LP=\sqrt{Q^2+U^2}$).  The Fe~{\sc i} 6842, 10783, 15534~\AA, \ and Ni~{\sc i} 6768~\AA \ lines are the weakest lines. Then, the Fe~{\sc i} 6173, and 6302~\AA \ spectral lines show similar signals that are always stronger than the previous signals. Lastly, another three lines show the strongest signals and correspond to Fe~{\sc i} 5250.2, 8468, and 15648~\AA. These results agree with the indices presented in Table~\ref{atomic_info}, indicating that we need to take the Land\'{e} factor, the line-to-continuum absorption ratio, and the wavelength range into which the transition falls in our selection of spectral lines into consideration \citep[see also the detailed discussion in][]{CabreraSolana2005}. In the case of the last three lines, we can expect linear polarisation signals around $1\times10^{-3}$ of $I_c$ for a reference horizontal magnetic field of 30~G.

The circular polarisation signals show a similar behaviour (bottom panel of Figure~\ref{LTE_pol}), although now the Fe~{\sc i} 6173~\AA \ transition is closer to the best three lines mentioned before. The infrared lines (light blue and red) reach the strong-field regime earlier than the others. However, their signals are already stronger than $1\times10^{-1}$ of $I_c$, so that this property will not affect their observability, even for noisy observations. Finally, all the studied lines generate circular polarisation signals stronger than $1\times10^{-3}$ of $I_c$ for a vertical field of 10~G.

\begin{figure*}
\begin{center} 
 \includegraphics[trim=0 0 0 0,width=18.2cm]{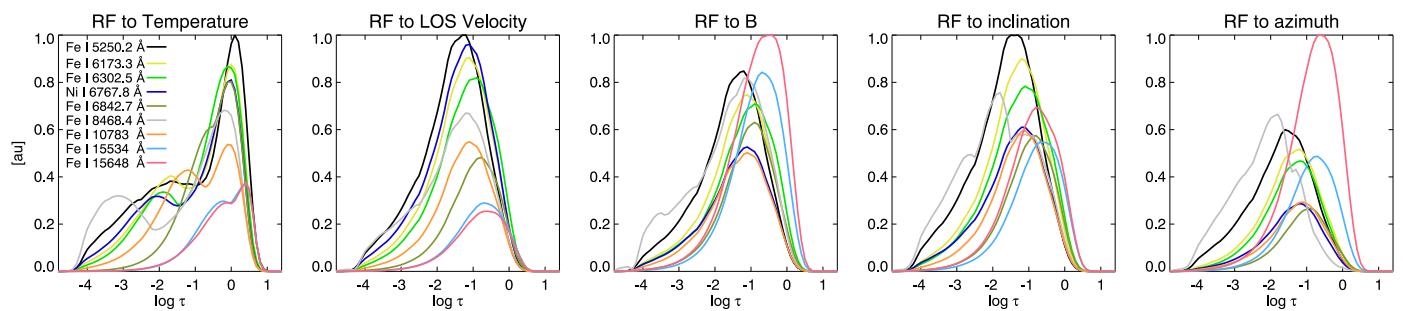}
 \caption{Maximum RF, from left to right, for the temperature, LOS velocity, and the three components of the magnetic field vector. Each line corresponds to the RF maximum value for the four Stokes parameters and all the computed wavelengths for the selected spectral lines (see colours and the legend in the leftmost panel). RFs are normalised to the maximum value of the nine spectral lines for each atmospheric parameter.}
 \label{LTE_1DRF}
 \end{center}
\end{figure*}

\subsection{Response functions to perturbations in the atmospheric parameters}

We computed the response functions \citep[RF, ][]{Landi1977} to perturbations in the atmospheric parameters for the nine most sensitive spectral lines studied before. We assumed LTE conditions, so that the SIR code can provide the analytical RFs for any given atmosphere. We used the HSRA model, adding a magnetic field of 500~G, and the inclination and azimuth were equal to 45 and 70 degrees, respectively.

To reduce the content of this work, we followed two approaches. We present the 2D ($\lambda-\log \tau$)\footnote{ We always refer to the continuum optical depth at 5000~\AA \ when using $\log \tau$.} Stokes $V$ RF to changes in the field strength and a more compact way of comparing the RF plotting its maximum values in a 1D plot. We show the first case in Figure~\ref{LTE_2DRF}, where all the spectral lines display an antisymmetric pattern. When the field strength is modified, the amplitude of the Stokes $V$ profile changes equally (in the absence of velocity asymmetries) for the negative and positive lobes. In terms of height, some spectral lines are sensitive to the magnetic field only at lower layers, that is, Fe~{\sc i} 6842, 10783, 15534, and 15648~\AA. The remaining spectral lines reach higher layers up to $\log \tau=-3.5$. Interestingly, the Fe~{\sc i} 15648~\AA \ transition appears to reach deeper layers than the other lines. As a side note, the infrared lines at 1.5 micron show multiple lobes because they already achieved the saturation regime with a field strength of 500~G. 

It is also useful to study the vertical stratification of the RF at specific wavelengths as was shown, for instance, by \citet{RuizCobo1992}, \citet{CabreraSolana2005}, and \citet{Fossum2005}. For this reason, we display in Figure~\ref{LTE_1DRF} a 1D plot computing the maximum of the RF functions to perturbations in different physical parameters.  In this case, we used the four Stokes parameters, so that we can compare the sensitivity of the spectral lines to the three components of the magnetic field vector. For each optical depth, we picked the highest value of the four Stokes parameters at any wavelength of the spectrum. 

Starting with the temperature, that is, with the leftmost panel, all the lines at lower layers contribute significantly. The Fe~{\sc i} 5250.2, 6173, 6302, and 8468~\AA \ and Ni~{\sc i} 6768~\AA \ lines extend to upper layers and reach up to $\log \tau=-4$. The LOS velocity RF shows a central peak for all the lines at about $\log \tau=-1.0,$  and again the Fe~{\sc i} 5250.2 and 8468~\AA \ lines reach highest in the atmosphere. The  Ni~{\sc i} 6768~\AA \ shows a high sensitivity to the LOS velocity as well, comparable to that of the Fe~{\sc i} 5250.2~\AA \ transition. A similar behaviour is found for the three components of the magnetic field, although in this case, the Ni transition is less sensitive than the two iron lines mentioned above. In the case of the field strength and azimuth, the dominant RF comes from Fe~{\sc i} 15648~\AA \ (red), and its peak is located at the bottom of the photosphere. This spectral line also reaches lower in the atmosphere for the three components of the magnetic field.

\section{Spectral line analysis based on 3D realistic simulations}

We highlight in Figure~\ref{context} the selected Field of View (FOV) with solid lines. We chose this region because it contains one of the enhanced network concentrations, but also weakly magnetised areas. In the following, we describe the height stratification of the atmospheric parameters in this region. However, before this, we explain why we transformed the original atmosphere from gas pressure and geometrical height to electronic pressure and optical depth. Ideally, the original parameters should be used because the simulation was run solving the equation of state, and therefore the relation between the atmospheric parameters is consistent for the whole range of heights. However, because we plan to compare the inversion results with the original atmosphere, we preferred to transform everything to optical depth values, which are required as input for SIR, and work from there.

\begin{figure*}
\begin{center} 
 \includegraphics[trim=0 0 0 0,width=18.2cm]{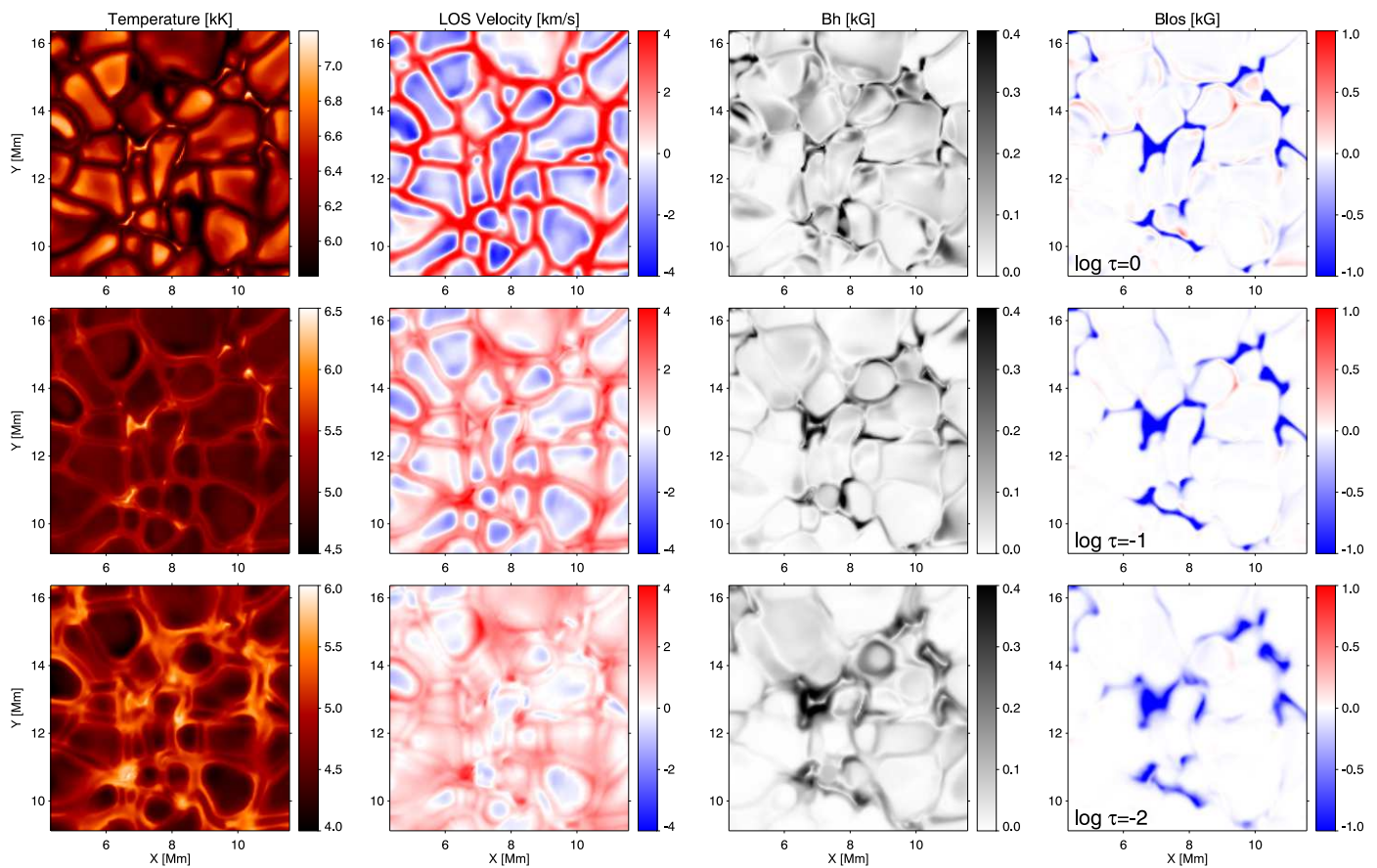}
 \caption{Snapshot 385 from the {\sc bifrost} enhanced network simulation. From left to right, we show temperature, LOS velocity, and the horizontal and longitudinal magnetic field. The rows display different atmospheric layers at $\log \tau=[0,-1,-2]$. The selected FOV corresponds to the region highlighted in Figure~\ref{context}.}
 \label{context_tau}
 \end{center}
\end{figure*}

\subsection{Bifrost-enhanced network simulation}

We used snapshot 385, which has been extensively employed in the past for a variety of studies. For example, it was used to characterise spectral lines or estimate the capabilities of the new generation of NLTE radiative transfer codes \citep[among others,][]{Stepan2016,delaCruzRodriguez2016, Sukhorukov2017, Bjorgen2018, Milic2018,daSilvaSantos2018,delaCruzRodriguez2019}. We show in Figure~\ref{context_tau} the spatial distribution of the temperature, LOS velocity, and the horizontal and longitudinal components of the magnetic field. We chose to display three optical depths (we always refer to the optical depth at 5000~\AA) at $\log \tau=[0,-1,-2]$  because the lines studied in this work are thought to form mostly in between these layers.

Starting with the temperature (leftmost column), the granulation pattern at lower heights with bright intergranular areas coincides with strong magnetic network concentrations. The upper layers display the reverse granulation pattern, with intergranules showing higher temperatures than granules, while the magnetic bright points are still hotter than their surroundings. The spatial distribution of the LOS velocities also shows the granulation pattern at lower heights, and then velocities that fade away at upper layers. In the case of the horizontal component of the magnetic field, we have narrow concentrations of more than 500~G that expand with height. These concentrations are in general located in the surroundings of the enhanced network patches. The longitudinal component is dominated by a single magnetic polarity that occupies broader areas at higher atmospheric layers. The longitudinal component can reach up to 2000~G in the low photosphere, although these values only correspond to the enhanced network areas. The rest of the FOV is characterised by weaker magnetic fields of less than 50~G.

\subsection{Spatial distribution of signals}

\begin{figure*}
\begin{center} 
 \includegraphics[trim=0 0 0 0,width=18.2cm]{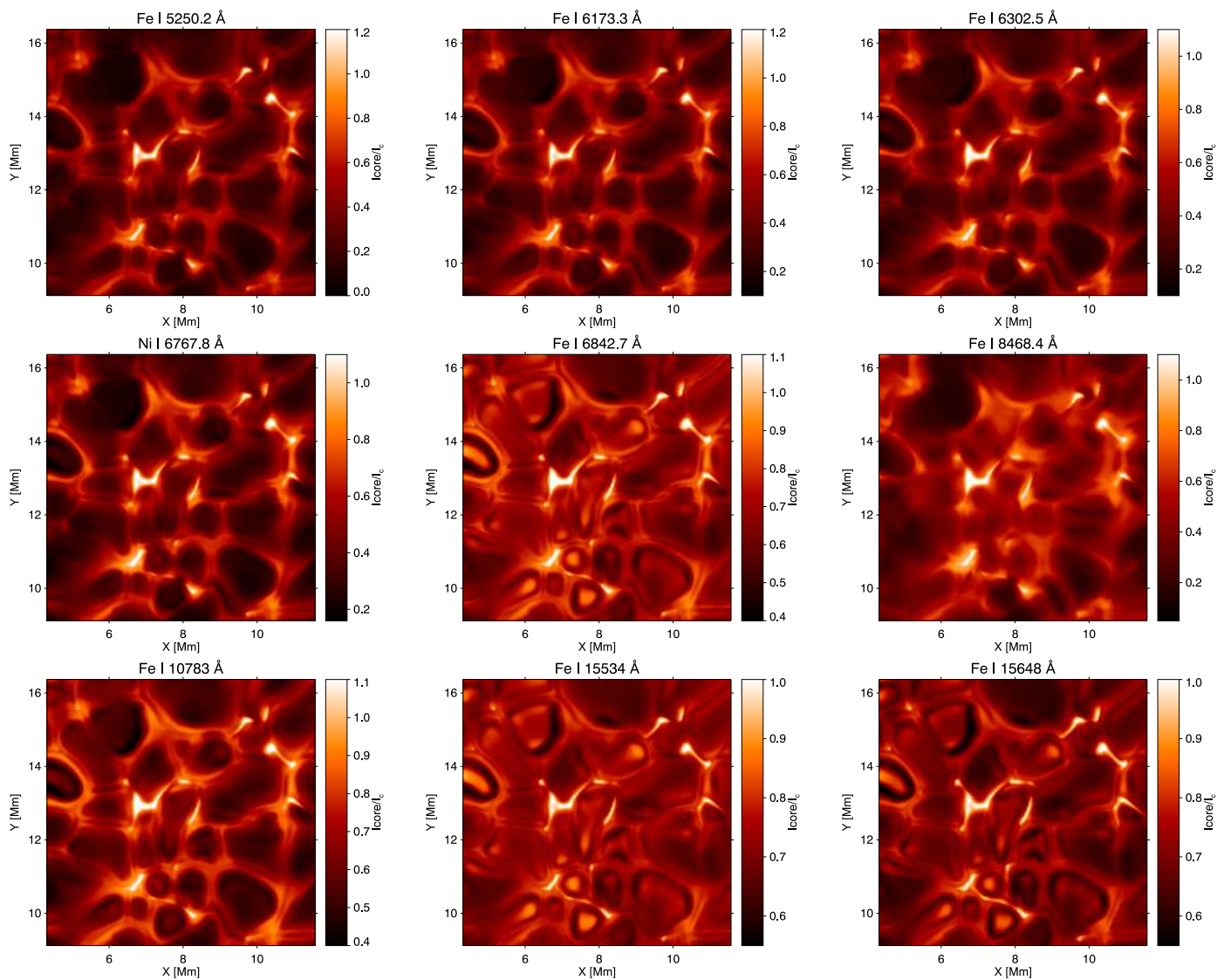}
 \caption{Spatial distribution of line core intensity signals for the lines of interest. The selected FOV corresponds to the highlighted area in Figure~\ref{context} that is described in Figure~\ref{context_tau}.}
 \label{Icore}
 \end{center}
\end{figure*}

\begin{figure*}
\begin{center} 
 \includegraphics[trim=0 0 0 0,width=18.2cm]{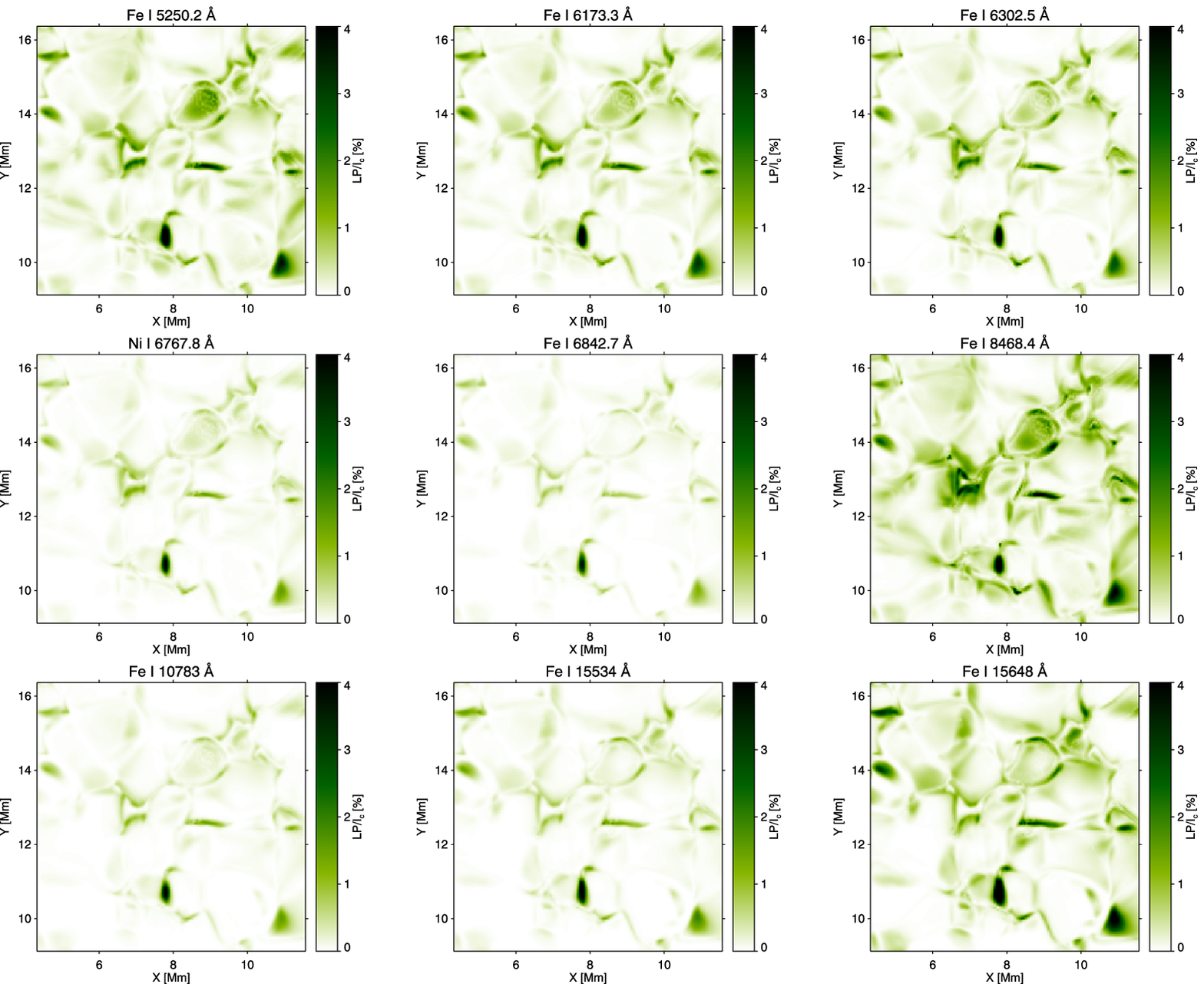}
 \caption{Spatial distribution of maximum linear polarisation signals for the lines of interest. White areas indicate a low degree of polarisation, and darker regions designate the highest signals. The selected FOV corresponds to the highlighted area in Figure~\ref{context} that is described in Figure~\ref{context_tau}.}
 \label{Ip}
 \end{center}
\end{figure*}

We show in Figure~\ref{Icore} the line core intensity of the nine selected spectral lines synthesised with SIR with the configuration explained in Section~\ref{method}. Some lines trace the temperature distribution at lower heights, partially resembling the granulation, such as Fe~{\sc i} 6842, 15534, and 15648~\AA. For all of them, we also see the bright features corresponding to the network patches. Then, the Fe~{\sc i} 5250, 6173, 6302, 8468, and 10783~\AA \ and Ni~{\sc i} 6767.8~\AA \ lines trace the temperature at higher photospheric layers, close to the pattern found at $\log \tau=-2$. The visible lines are very similar between them, while Fe~{\sc i}~10783~\AA \ shows more defined structures that would indicate a lower height of formation. The infrared Fe~{\sc i}~8468~\AA \, shows the opposite behaviour, with a more diffused pattern that could be associated with a higher formation height, similar to what we found for the RF analysis.

We present in Figure~\ref{Ip} the spatial distribution of maximum linear polarisation signals. We saturated the colour scale to 4~$\%$ of the local continuum intensity, that is, 0.04 of $I_c$. Weaker lines such as Fe~{\sc i} 6842, 10783, and 15534~\AA \ and Ni~{\sc i} 6767.8~\AA \ do not reach the saturation value for most of the selected field of view, indicating that they are less suitable for observations with low magnetic activity. On the other hand, Fe~{\sc i} 5250, 6173, 6302, 8468, and 15648~\AA \ reach the bar scale saturation in various areas, for instance at the edges of the network patches. Of the five lines, Fe~{\sc i} 5250 and 8468~\AA \ are similar in terms of spatial distribution and the amplitude of the signals, with the latter showing the highest amplitude. They also display linear polarisation signals inside the large granule at [9,14]~Mm. These signals correspond to the horizontal magnetic field at higher layers, see $\log \tau=-2$ in Figure~\ref{context_tau}. This explains why they are not present in the Fe~{\sc i} 15648~\AA \ line. However, the infrared line shows stronger polarisation signals in different locations such as [8,11] or [5.3,15.7]~Mm.

\begin{figure*}
\begin{center} 
 \includegraphics[trim=0 0 0 0,width=18.0cm]{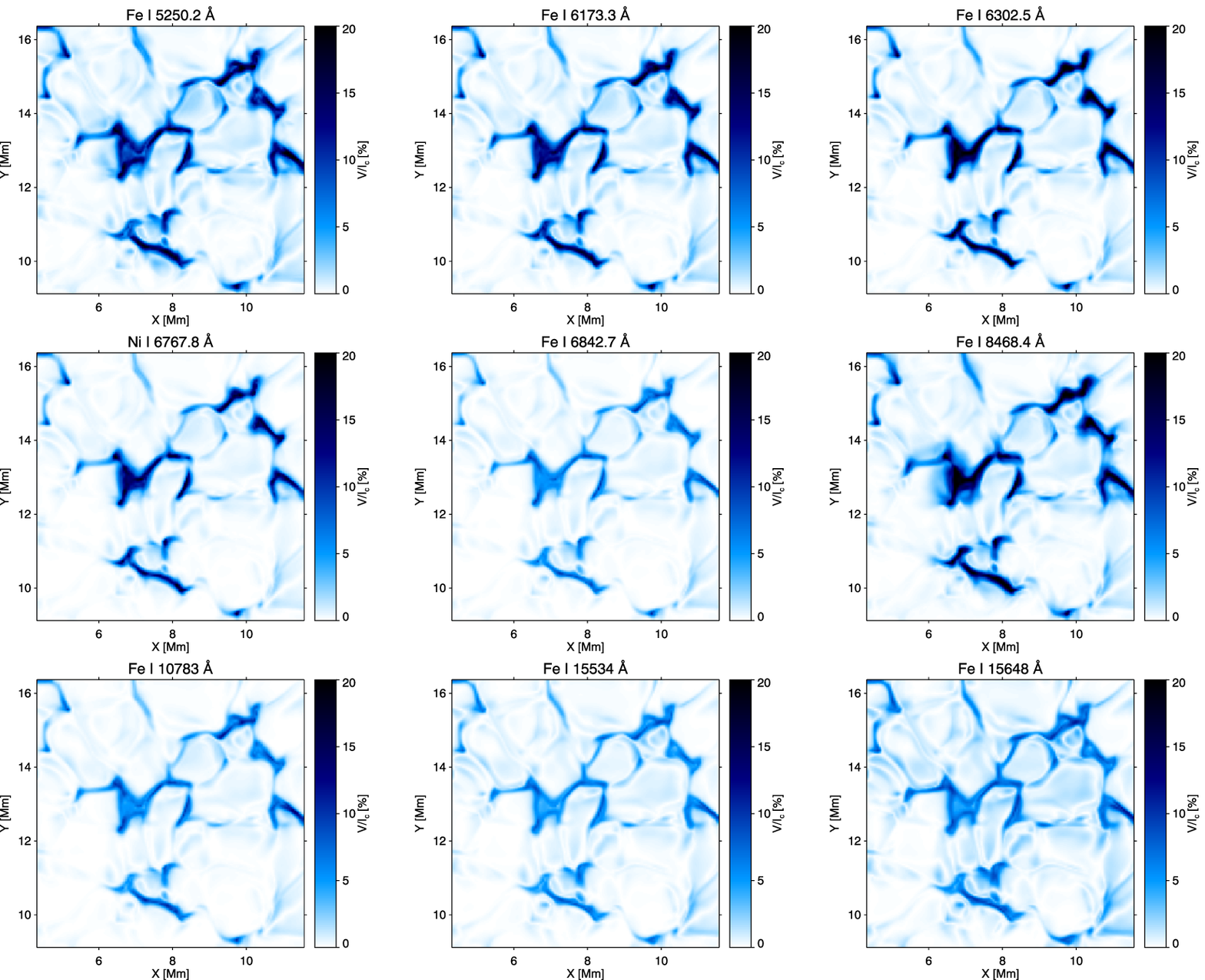}
 \caption{Spatial distribution of maximum circular polarisation signals for the lines of interest. White areas indicate a low polarisation degree, and darker regions designate the largest amplitude. The selected FOV corresponds to the highlighted area in Figure~\ref{context} that is described in Figure~\ref{context_tau}.}
 \label{vmax}
 \end{center}
\end{figure*}

The results for the maximum circular polarisation signals are shown in Figure~\ref{vmax}. In this case, we saturated the colour bar to $20~\%$ of the local continuum intensity,  that is, 0.20 of $I_c$. The Fe~{\sc i} 6842, 10783, and 15534~\AA \ transitions are again weaker than the others, and they never reach the colour saturation. However, in this case, the Fe~{\sc i} 15648~\AA \ line also shows a similar spatial distribution because the infrared transition reaches the polarisation saturation regime (see also Figure~\ref{LTE_pol}) earlier than the lines in the visible. The Stokes profiles are entirely split, that is, the three Zeeman components are resolved, since about 500~G. In other words, magnetic fields stronger than the mentioned value will increase the separation of the two $\sigma$ components, while the Stokes $V$ amplitude remains almost constant. This behaviour is partially detected for the visible Fe~{\sc i} 5250~\AA \ line but only in the most active magnetic patch at [6.3,12.7]~Mm. For the remaining FOV, the spatial distribution of signals is similar for the Fe~{\sc i} 5250, 6173, 6302, and 8468~\AA \ spectral lines. The latter seems to be sensitive to slightly higher layers, with the polarisation signals occupying more extensive areas. Finally, we also find that the Ni~{\sc i} 6767.8~\AA \ transition shows significant Stokes~$V$ signals, comparable to those of Fe~{\sc i} 6302~\AA, for example, confirming that it is a good candidate for measuring longitudinal fields.

\begin{figure*}
\begin{center} 
 \includegraphics[trim=0 0 0 0,width=18.2cm]{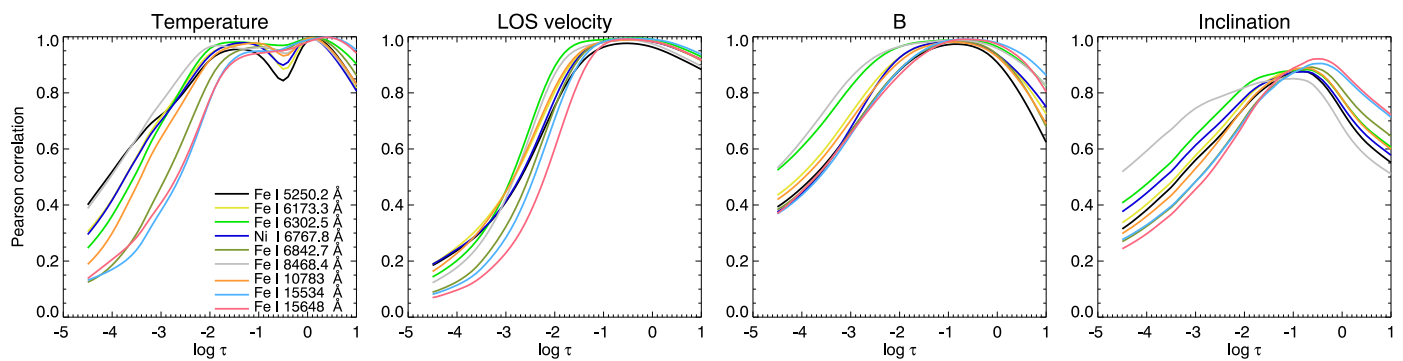}
 \caption{Spatial correlation between the input atmosphere and the atmospheric parameters obtained from the inversion. From left to right, we show the temperature, LOS velocity, field strength, and inclination. We computed the correlation between the input and the inferred atmospheric parameters at different optical depths.}
 \label{inver}
 \end{center}
\end{figure*}

\begin{figure*}
\begin{center} 
 \includegraphics[trim=0 0 0 0,width=18.2cm]{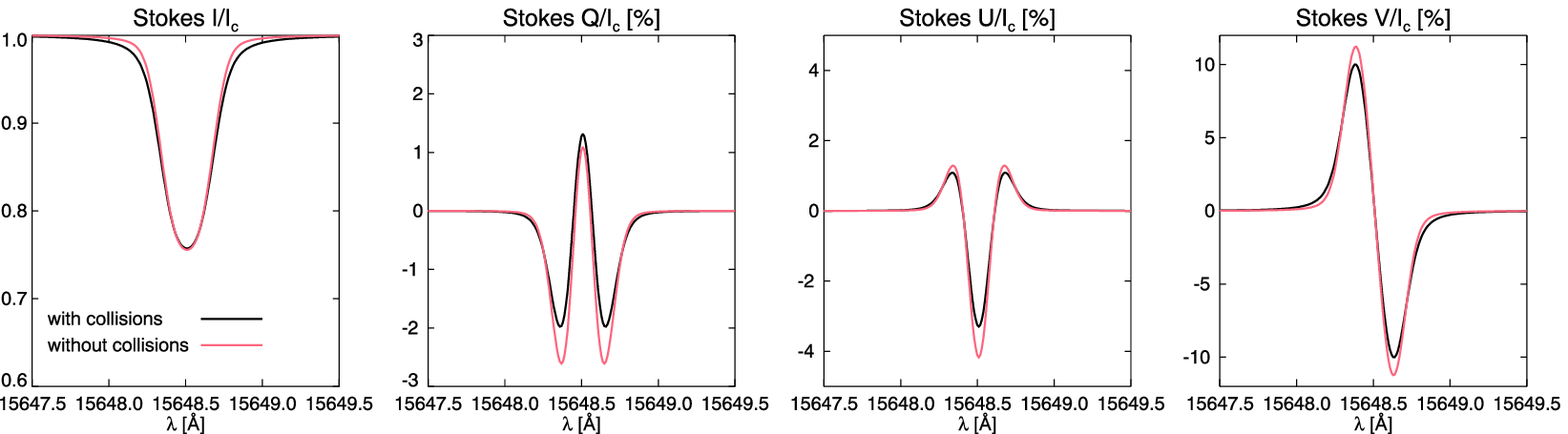}
 \caption{Comparison between the synthetic profiles for the Fe~{\sc i} 15648~\AA \ transition when using the theory of \cite{Anstee1995} for collisions with neutral hydrogen (black) and the case where no broadening process is included (red). We use the HSRA atmosphere with a constant magnetic field strength of 300~G, 45 degrees of inclination, and 70 degrees of azimuth.}
 \label{Prof_barklem}
 \end{center}
\end{figure*}

\subsection{Inversions of the Stokes profiles}\label{Inverconfig}

We wish to improve our comparison of photospheric lines running inversions of the full Stokes vector. The aim is to estimate how much information we recover when we fit synthetic spectra. We used the profiles from the FOV highlighted in Figure~\ref{context}, and we inverted them with the SIR code. We fitted each spectral line separately because we aimed to compare how much information we extracted from each one. We are not interested at this moment in simulating a particular observation, therefore we opted not to include any instrumental degradation or noise.

The inversion was made with SIR using a single magnetic component parametrised with five nodes in temperature, three for the LOS velocity, magnetic field strength, and inclination, and two nodes for the magnetic field azimuth. We did not invert the microturbulence or the macroturbulence, which are both null and equal to the value used for the synthesis. We ran the inversion in parallel using the method described in Section 5 of  \citet{Gafeira2021}, initialising each pixel with three different atmospheres. All of them are modifications of the HSRA model we used in previous sections with different magnetic field configurations. After the three runs, the code automatically picked the solution that produced a better $\chi^{2}$ value.

The inversion process provides the stratification with height of various atmospheric parameters. This information is obtained for each pixel of the FOV highlighted in Figure~\ref{context}. This 3D information is inferred for each spectral line of interest, a total of nine transitions. This means that comparing the results would require multiple figures with a format similar to Figures~\ref{Icore}, \ref{Ip}, and \ref{vmax}. We therefore decided to choose a different approach in this work. We computed the correlation of the original and the inferred atmosphere for a given atmospheric parameter at a given optical depth over the entire FOV (150$\times$150 pixels) highlighted in Figure~\ref{context}. We computed the correlation for all the optical depths that we have in the inverted atmosphere, obtaining an estimate of how close to the original atmosphere is to the inverted one on average over the complete FOV. Thus, we simplified the visualisation to a single height-dependent vector, that is, the correlation, for each atmospheric parameter and each spectral line.

We present in Figure~\ref{inver} the results of the correlation between the atmospheric parameters over the entire FOV for each spectral line using the same colour code as in Figure~\ref{LTE_1DRF}.  Starting with the temperature (leftmost panel), we can detect a high correlation (almost 1) in a wide range of heights, $\log \tau=[0,-2.0]$, for most of the transitions. The correlation at the deepest layers is also higher for the spectral lines above 1 micron, while the upper layers show the opposite behaviour with the visible and red transitions, obtaining a high correlation up to $\log \tau \sim -3$. Continuing with the LOS velocity (second panel from the left), all the candidates achieve high correlation values in the range of optical depths $\log \tau=[0,-1.5],$ indicating that the velocity stratification is covered at these heights with any of the spectral lines.  Again, the infrared transitions above one micron show a high correlation at deeper layers that drops to low values at slightly lower layers than those of the visible lines. The situation is similar for the magnetic field strength, where the correlation in the range of optical depths $\log \tau=[0,-2.0]$ is high. However, in this case, it appears to be more evident that the iron lines in the visible and red part of the spectrum, 6302~\AA \ and 8468~\AA ,\ show a high correlation at moderately high layers. The situation for the magnetic field inclination is similar, where the correlation drops faster for the infrared (above 1 micron) spectral lines. However, it is also true that these spectral lines reach the highest correlation (at any height), which again indicates a high sensitivity to the transversal component of the magnetic field.

\begin{figure*}
\begin{center} 
 \includegraphics[trim=0 0 0 0,width=18.2cm]{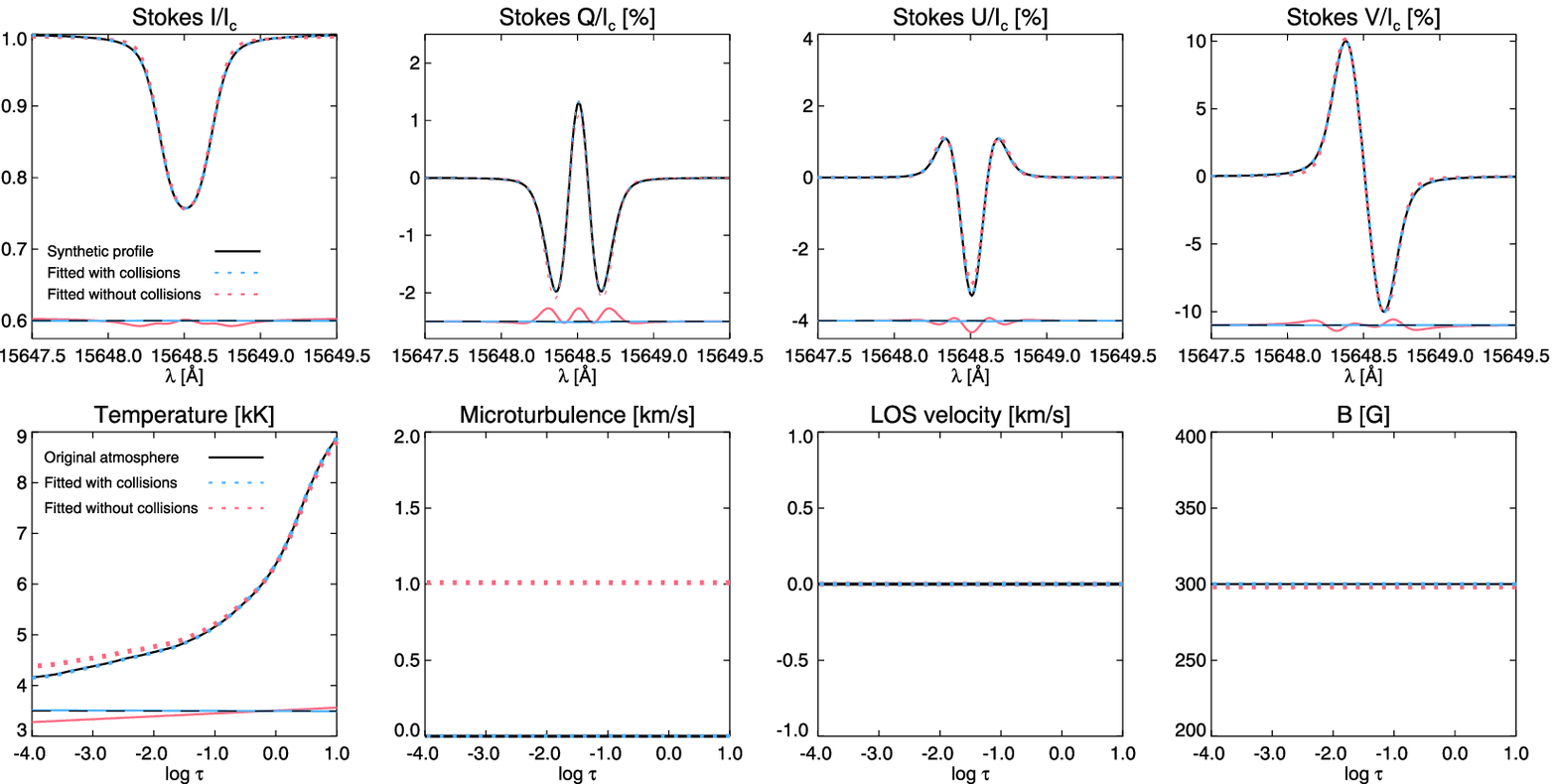}
 \caption{Results for the inversion of the Fe~{\sc i} 15648~\AA \ synthetic profile presented in Fig~\ref{Prof_barklem} in black (top row). This profile was synthesised considering collisions with H. The effects of fitting this profile taking into account collisions with neutral hydrogen are plotted in blue, and those without considering collisions are shown in red. The bottom row displays the atmosphere that produced the input profiles (black) and the atmosphere inferred considering collisions with H (blue) and considering no collisions (red). We show the differences between the two studies at the bottom of the four Stokes parameters and the temperature panels.}
 \label{Prof_barklem2}
 \end{center}
\end{figure*}

\section{Discussion}

\subsection{Effect of collisions with neutral hydrogen on the inversions of spectropolarimetric data}

We assumed so far that the theory presented in \cite{Anstee1995} should be included in the future when spectropolarimetric observations of the lines compared here are inverted. For this reason, we focused only on computing or updating the cross-section values for these transitions. However, it is worth showing the change in Stokes profiles when we include or exclude collisions with neutral hydrogen. Most importantly, however, we also wish to show the effect on the inferred atmospheric parameters when collisions are included or excluded during the inversion process.

We start with the effect on the polarimetric signals when the theory of \cite{Anstee1995} is assumed. We used the original HSRA atmosphere, adding a constant magnetic field strength of 300 G, 45 degrees of inclination, and 70 degrees of azimuth. We show in Figure~\ref{Prof_barklem} the results of the synthesis of the Fe~{\sc i} 15648~\AA \ transition, chosen as a reference spectral line for this discussion, when collisions with neutral hydrogen are included (black) or excluded (red). The intensity profile computed without collisions is narrower and slightly deeper, which produces polarimetric profiles with a larger amplitude. These differences will affect the inferred atmospheric parameters, and we wish to quantify which parameters are more affected.

We performed two inversions of the Stokes profiles synthesised with collisions (black in Fig~\ref{Prof_barklem}). In one case, we provided the code with the cross-section parameters, and the second case was without collisions. The results of the fit are shown in the top row of Figure~\ref{Prof_barklem2} with the original Stokes parameters in black, the results of the fit including collisions in blue, and in red that without considering collisions. The two fits are almost perfect; the one without collisions is less accurate. In the bottom row, we show the inferred atmospheric parameters using the same colour code. When collisions are included (blue), the fits match the original atmosphere very well. However, when the theory of \cite{Anstee1995} is not included in the fitting, the code has to reproduce the larger broadening of the intensity profile modifying the thermal properties of the atmosphere (see red). It raises the temperature above $\log \tau\sim-0.7$ and also adds a microturbulence enhancement of 1~km/s. For the remaining physical parameters (we only show the LOS velocity and the magnetic field strength), the atmospheric values are identical for both inversions. These results indicate that if we do not consider the broadening due to collisions with neutral hydrogen when fitting observations, we modify the temperature and add an artificial microturbulence that will make our results less accurate.

\subsection{Suitability of the LTE approximation for photospheric lines}

We concentrated our efforts on comparing multiple spectral lines at identical conditions. These conditions mean solving the radiative transfer equation for the four Stokes parameters under the LTE approximation, which is widely used to analyse spectropolarimetric observations. However, several studies have found deviations from LTE for the photospheric lines presented here. The first work we are aware of is from \citet{Athay1972}, but there are additional ones from (among others, \cite{Rutten1982}, \cite{Rutten1988}, \cite{Shchukina2001} or more recent publications including 3D radiative transfer computations \citep{Holzreuter2012,Holzreuter2013,Holzreuter2015,Smitha2020}. We used the LTE approximation in this work because we aimed to compare a vast number of transitions, and LTE is the best trade-off between accuracy and computational time. However, we plan to solve the radiative transfer equation for the transitions of interest in NLTE in future publications.  

\section{Summary}

This work is the first step in a series of publications in which we plan to revise the literature searching for spectral line candidates, and we compare them using modern tools such as 3D realistic simulations. We dedicated some effort to extracting the atomic information of these lines and to computing the cross-section parameters due to collisions with H. The second step is crucial to avoid mixing the line broadening induced by the temperature or plasma flows with that intrinsic of the atomic transition. Then, we compared at identical conditions the nine best candidate photospheric lines. We examined their properties using the HSRA atmosphere first, computing their maximum polarisation signals and their response to perturbations in the atmospheric parameters. The outcome of these studies indicated that four spectral lines stand out, Fe~{\sc i} 5250, 6302, 8468, and 15648~\AA. They produce higher polarisation signals and with a broader height coverage. We then employed 3D numerical simulations to examine the spatial distribution of line core intensity and maximum polarisation signals. The results are similar to the previous study, with the four lines showing the largest amplitude. The Fe~{\sc i}  15648~\AA \ line forms lowest in the atmosphere, while Fe~{\sc i} 5250, 6302, and 8468~\AA \ form at similar heights and above the previous line. We also included the results from the inversions of the synthetic data. The results confirmed what we found before, showing a good correlation with the input atmosphere in deeper layers for Fe~{\sc i} 15648~\AA \ and a better correlation at higher layers for the other three transitions.

\section{Recipe for optimised measurements of photospheric magnetic fields?}

We started this work with the aim to offer a recipe for the best spectral line for photospheric spectropolarimetry. However,  creating such a recipe could seem like a challenge in view of the infinite options offered by the DKIST and EST telescopes. Most importantly, we admit that there is no definitive answer regarding a spectral line that can do everything, and the perfect recipe probably always requires more than just one ingredient. We therefore have four lines that outshine the rest in quiet-Sun conditions: Fe~{\sc i} 5250, 6302, 8468, and 15648~\AA. They produce the strongest signals, show the largest RFs, and cover the broadest range of heights. The 15648~\AA \ line forms lower in the atmosphere. The opposite scenario holds for the other three; they extend higher in the photosphere but are less sensitive to deeper layers. The best recipe therefore appears to imply combining at least one of the first three spectral lines with the Fe~{\sc i} 15648~\AA \ to maximise our sensitivity to the atmospheric parameters in the broadest possible height range. It is exciting that the Fe~{\sc i} 5250, 6302 and 8468~\AA \ lines are more than 3000 (2000)~\AA \ apart, which means that we can choose one instead of the other based on what we wish to observe in a given spectral range. At the same time, we do not wish to leave the Fe~{\sc i} 6173~\AA \ transition out of the formula. It is a slightly worse than the other three visible-red transitions, but also has particular advantages. It is an isolated line with a clean continuum that makes it useful, for example, for space observations. Moreover,  we should not exclude the Ni~{\sc i} 6767.8~\AA \ spectral line from the recipe. It was almost the most sensitive spectral line when we computed the RF to the LOS velocity, and although it produced low linear polarisation signals in the Bifrost simulation, the Stokes~$V$ signals were comparable to those generated by the best spectral lines.

We strongly recommend combining any or several of these spectral lines with the Fe~{\sc i} 15648~\AA \ because it reaches lower than any of those transitions. It will also excel in quiet-Sun observations where the magnetic field is weakest. 

\section*{Acknowledgements}
C. Quintero Noda was supported by the EST Project Office, funded by the Canary Islands Government (file SD 17/01) under a direct grant awarded to the IAC on ground of public interest, and this activity has also received funding from the European Union's Horizon 2020 research and innovation programme under grant agreement No 739500. This work was supported by the Research Council of Norway through its Centres of Excellence scheme, project number 262622, and through grants of computing time from the Programme for Supercomputing. P. S. Barklem received financial support from the Swedish Research Council. The Centre for Earth and Space Research of University of Coimbra is funded by National Funds through FCT - Foundation for Science and Technology (project: UID/MULTI/00611/2019) and FEDER – European Regional Development Fund through COMPETE 2020 – Operational Programme Competitiveness and Internationalization (project: POCI-01-0145-FEDER-006922)  This work has been supported by the Spanish Ministry of Economy and Competitiveness through the projects ESP-2016-77548-C5-1-R and by Spanish Science Ministry ``Centro de Excelencia Severo Ochoa'' Program under the grant SEV-2017-0709 and project RTI2018-096886-B-C51. D. Orozco Su\'{a}rez also acknowledges financial support through the Ram\'{o}n y Cajal fellowships. This work was supported by the ISAS/JAXA Small Mission-of-Opportunity program for novel solar observations and the JSPS KAKENHI Grant Number JP18H05234.

\bibliographystyle{aa} 
\bibliography{ltelines} 

\begin{appendix}

\section{ Fe~{\sc i} 15652.874~\AA \ }

The pair of Fe~{\sc i} infrared lines at 15648 and 15652~\AA \  has been extensively used in the literature for polarimetric measurements since the early 1990s. Among many other works, we have the series of works entitled \textit{Infrared lines as probes of solar magnetic features} that started with \cite{Muglach1992}, or the publications of, for instance, \citet{Mathew2003}, \citet{Khomenko2005}, \citet{BelloGonzalez2013}, and \citet{Beck2017}. If we examined the atomic information for both lines, presented in Table~\ref{cross}, the second line cannot be treated assuming LS-coupling. We therefore excluded it from Table~\ref{atomic_info} in Section 2. However, we plan to describe how it can be appropriately included in the SIR code for future inversions. The same modifications can also probably be implemented in any of the new codes, such as SNAPI \citep{Milic2018}, StiC \citep{delaCruzRodriguez2019}, or FIRTEZ-dz \citep{PastorYabar2019}.

The effective Land\'{e} factor of the Fe~{\sc i} 15652~\AA \ transition  should be evaluated considering a $J-K$ coupling \citep[also known as $J_1$-$l$ coupling, see ][]{Landi2004}. Paraphrasing the authors, we have that another case where it is possible to find an analytical expression for the Land\'e factor is the case of the so-called $J_1 -l$ coupling. In this coupling scheme, a \textit{parent} level of orbital angular momentum $L_1$ and spin $S_1$ couples its total angular momentum $J_1$ with the orbital angular momentum $l$ of a further electron, to give an angular momentum $K$ which in its turn couples with the electron spin to give the total angular momentum $J$. To evaluate the Land\'e factor for this type of transition, we therefore need to provide the quantum numbers $L_1$, $S_1$ , and $J_1$ of the parent level, the quantum numbers of the last electron $l$ and $s=1/2$, the orbital angular momentum of the coupling $K$, and the total angular momentum $J$.

In the case of SIR, we need the spectroscopic information $7D$ $5.0-(6D4.5)f2k$ 4. The first is the LS-coupling description of the lower level and the second is the $J_1$-$l$ coupling information of the upper level. Every time the code finds a parenthesis in the atomic information, as for the upper level, it reads the information inside as the parent term, and it considers that the $J-K$ coupling is defined. In this example, the configuration of the upper level says $(6D4.5)f2k$ 4, which means that $S_1=6$, $L_1=2$ (because the angular momentum is $D$), $J_1=4.5$, $l=3$ (because the angular momentum of the last electron is $f$), $s=1/2$ (we do not need to provide it because the spin of the additional electron is always taken as 1/2). Then, the orbital angular momentum of the coupling $K$ is given by the NIST database as $K=7/2$. However, in SIR, following the previous convention, we use characters, similar to the linear angular momentum, with $p=0.5$, $f=1.5$, $ h=2.5$, $k=3.5$, $m=4.5$, $o=5.5$, $r=6.5$, $t=7.5$, $u=8.5$, $v=8.5$, and $w=9.5$. This means that in the input file for the code, we write $(6D4.5)f2k$.  Lastly, the total angular momentum is $J=4$.

Gathering all this information, we can use the following equation \citep[see ][]{Landi2004}:

\begin{equation}
g_{J_1l}= 1+ \gamma(J,1/2,K) + \gamma(J,K,1/2)\gamma(K,J_1,l)\gamma(J_1,S_1,L_1)
,\end{equation}
where $\gamma$ is defined as
\begin{equation}
\gamma(A,B,C)=\frac{A(A+1)+B(B+1)-C(C+1)}{2A(A+1)}
.\end{equation}
For the Fe~{\sc i} 15652~\AA \  infrared transition, we have then that $g_{J_1l} = 2.097$. We modified the SIR code to accept any transition in $LS$ or $J-K$ coupling.\ Anyone interested in using the new version or knowing more about how we did it is more than welcome to contact us.

\end{appendix}

%
%

%
%
\end{document}